\documentclass[12pt]{article}
\usepackage{amsmath}
\usepackage{pstricks}
\usepackage{tikz-cd}

\usepackage{youngtab}
\usepackage{bbm}
\usepackage{slashed}
\usepackage{amssymb}
\def\half{\frac{1}{2}}

\newfont{\bbbold}{msbm10 scaled \magstep1}

\def\bbC{\mbox{\bbbold C}}

\def\bbF{\mbox{\bbbold F}}

\def\bbR{\mbox{\bbbold R}}

\def\cA{{\cal A}}

\def\cD{{\cal D}}

\def\cF{{\cal F}}

\def\cL{{\cal L}}

\def\cN{{\cal N}}
\def\cO{{\cal O}}

\def\cR{{\cal R}}
\def\cS{{\cal S}}
\def\cT{{\cal T}}

\newfont{\goth}{eufm10 scaled \magstep1}

\def\gc{\mbox{\goth c}}

\def\gg{\mbox{\goth g}}
\def\gh{\mbox{\goth h}}
\def\gi{\mbox{\goth i}}

\def\gn{\mbox{\goth n}}
\def\go{\mbox{\goth o}}
\def\gp{\mbox{\goth p}}

\def\gs{\mbox{\goth s}}

\def\gu{\mbox{\goth u}}

\def\a{\alpha}\def\adt{\dot \alpha}
\def\b{\beta}\def\bdt{\dot \beta}
\def\c{\gamma}\def\C{\Gamma}\def\cdt{\dot\gamma}
\def\d{\delta}\def\D{\Delta}\def\ddt{\dot\delta}
\def\ve{\varepsilon}
\def\vf{\varphi}
\def\h{\eta}

\def\k{\kappa}
\def\l{\lambda}

\def\r{\rho}
\def\s{\sigma}

\def\be{\begin{equation}}\def\ee{\end{equation}}
\def\bea{\begin{eqnarray}}\def\eea{\end{eqnarray}}
\def\barr{\begin{array}}\def\earr{\end{array}}

\def\o{\omega}\def\O{\Omega}
\def\U{\Upsilon}
\def\del{\partial}

\def\xz{\times}

\def\gspin{\gs\gp\gi\gn}
\let\la=\label

\def\nn{\nonumber}
\def\bd{\begin{document}}
\def\ed{\end{document}}
\def\ba{\begin{array}}
\def\ea{\end{array}}
\def\bea{\begin{eqnarray}}
\def\eea{\end{eqnarray}}
\def\ft#1#2{\tfrac{#1}{#2}}
\def\fft#1#2{\frac{#1}{#2}}
\def\sst#1{{\scriptscriptstyle #1}}
\def\oneone{\rlap 1\mkern4mu{\rm l}}

\newcommand{\eq}[1]{(\ref{#1})}
\newcommand{\w}[1]{\\[0.#1cm]}
\def\eqs#1#2{(\ref{#1}-\ref{#2})}
\def\det{{\rm det\,}}
\def\tr{{\rm tr}}
\def\ad{{\rm ad}}

\newcommand{\hoch}[1]{$\, ^{#1}$}
\newcommand{\imperial}{\it\small Theoretical Physics Group, Imperial College London\\ Prince Consort Road, London SW7 2AZ, UK}
\newcommand{\kings}
{\it\small Department of Mathematics, King's College, University of London\\ Strand, London WC2R 2LS, UK}
\newcommand{\uu}
{\it\small Department of Theoretical Physics, Uppsala, Sweden}
\newcommand{\hip}
{\it\small HIP-Helsinki Institute of Physics, P.O. Box 64 FIN-00014
University of Helsinki, Suomi-Finland}
\newcommand{\stock}
{\it\small Department of Theoretical Physics, Stockholm, Sweden}
\newcommand{\golm}
{\it\small AEI, Max Planck Institut f\"ur Gravitationsphysik\\ Am M\"{u}hlenberg 1, D-14476 Potsdam, Germany}
\makeatletter
\renewcommand\theequation{\thesection.\arabic{equation}}
\@addtoreset{equation}{section} \makeatother

\newcommand{\sa}{/ \hspace{-1.2ex}}
\newcommand{\saa}{/ \hspace{-1.4ex}}
\newcommand{\saaa}{\, / \hspace{-1.6ex}}
\newcommand{\Scal}[1]{\Bigl ({#1} \Bigr )}
\newcommand{\scal}[1]{\bigl ({#1} \bigr )}
\newcommand{\rd}[1]{{\color{red}{#1}}}

\newcommand{\CR}{\nonumber \\*}

\newcommand{\trace}{\hbox {tr}~}
\newcommand{\traceS}{\hbox {tr}_{\scriptscriptstyle \mathfrak{S}}~}

\DeclareMathAlphabet{\mathpzc}{OT1}{pzc}{m}{it}
\def\BRST{\,\mathpzc{s}\,}
\def\aBRST{{\scriptstyle (\mathpzc{s})}}
\def\q{{{\scriptscriptstyle (Q)}}}
\def\qs{{\scriptscriptstyle (Q\mathpzc{s})}}
\def\Qsla{{\mathcal{S}_{\q}}}
\def\Slav{{\mathcal{S}_\aBRST}}
\def\epsilonb{{\overline{\epsilon}}}
\def\bulletup{{\scriptstyle \bullet}}

\newcommand{\gra}[2]{{\scriptscriptstyle (#1 , #2 )}}
\newcommand{\ord}[1]{{\scriptscriptstyle (#1)}}

\def\cL{{\cal L}}
\def\cN{\mathcal{N}}
\def\cO{\mathcal{O}}

\def\ie{{\it i.e.}\ }
\def\eg{{\it e.g.}\ }

\newcommand{\sfrac}[2]{{\scriptstyle \frac{#1}{#2}}}
\newcommand{\stfrac}[2]{{\scriptscriptstyle \frac{#1}{#2}}}

 \def\balpha{{\overline{\alpha}}}
 \def\bbeta{{\overline{\beta}}}
 \def\bgamma{{\overline{\gamma}}}
 \def\bdelta{{\overline{\delta}}}
 \def\bepsilon{{\overline{\epsilon}}}
 \def\bvarepsilon{{\overline{\varepsilon}}}
 \def\bzeta{{\overline{\zeta}}}
 \def\bareta{{\overline{\eta}}}
 \def\btheta{{\overline{\theta}}}
 \def\bvartheta{{\overline{\vartheta}}}
 \def\biota{{\overline{\iota}}}
 \def\bkappa{{\overline{\kappa}}}
 \def\blambda{{\overline{\lambda}}}
 \def\bmu{{\overline{\mu}}}
 \def\bnu{{\overline{\nu}}}
 \def\bxi{{\overline{\xi}}}
 \def\bpi{{\overline{\pi}}}
 \def\brho{{\overline{\rho}}}
 \def\bvarrho{{\overline{\varrho}}}
 \def\bsigma{{\overline{\sigma}}}
 \def\bvarsigma{{\overline{\varsigma}}}
 \def\btau{{\overline{\tau}}}
 \def\bphi{{\overline{\phi}}}
 \def\bvarphi{{\overline{\varphi}}}
 \def\bchi{{\overline{\chi}}}
 \def\bpsi{{\overline{\psi}}}
 \def\bomega{{\overline{\omega}}}

\newcommand{\ihalf}{{\textstyle{\frac i 2}}}
\newcommand{\quart}{{\textstyle{\frac14}}}

\def\thalf{{\textrm{\tiny\textonehalf}}}
\def\tquarter{{\textrm{\tiny\textonequarter}}}
\def\Ko{{\scriptscriptstyle K}}
\def\tKo{\scriptscriptstyle k }
\def\corr{$\clubsuit$}

\newcommand{\auth}{\large P.S.\ Howe${}^{a,}$\footnote{email: paul.howe@kcl.ac.uk} and U. Lindstr\"om${}^{b,c,}$\footnote{email: ulf.lindstrom@physics.uu.se}}

\begin{document}

\renewcommand{\thefootnote}{\fnsymbol{footnote}}

\null
\begin{flushright}
{\small UUITP-49/20}\\
\vskip 1.5 cm
\end{flushright}

\begin{center}
{\Large{\bf Superconformal geometries {and local twistors}}}
\vspace{.75cm}

\auth
\end{center}
\vspace{.5cm}
\centerline{${}^a${\it \small Department of Mathematics, King's College London}}
\centerline{{\it \small The Strand, London WC2R 2LS, UK}}
\vspace{.5cm}
\centerline{${}^b${\it \small Department of Physics, Faculty of Arts and Sciences,}}
\centerline{{\it \small Middle East Technical University, 06800, Ankara, Turkey}}
\vspace{.5cm}
\centerline{${}^c${\it \small Department of Physics and Astronomy, Theoretical Physics, Uppsala University}}
\centerline{{\it \small SE-751 20 Uppsala, Sweden }}

\vspace{1cm}


\centerline{{\bf Abstract}}
\vskip .5cm
 Superconformal geometries in spacetime dimensions $D=3,4,{5}$ and $6$ are discussed in terms of local supertwistor bundles over standard superspace. These natually admit superconformal connections as matrix-valued one-forms. In order to make contact with the standard superspace formalism it is shown that one can always choose gauges in which the scale parts of the connection and curvature vanish, in which case the conformal and $S$-supersymmetry transformations become subsumed into super-Weyl transformations. The number of component fields can be reduced to those of the minimal off-shell conformal supergravity multiplets by imposing constraints which in most cases simply consists of taking the even covariant torsion two-form to vanish. This must be supplemented by further dimension-one constraints for the maximal cases in $D=3,4$. The subject is also discussed from a minimal point of view in which only the dimension-zero torsion is introduced. Finally, we introduce a new class of supermanifolds, local super Grassmannians, which provide an alternative setting for superconformal theories.

\vspace{0.5cm}
\small
\begin{center}
{Dedicated to this year's Nobel Laureate in Physics, Sir Roger Penrose,\\ in
appreciation of his many achievements, including the invention of twistor theory}
 \end{center}
 \normalsize

\renewcommand{\thefootnote}{\arabic{footnote}}
\setcounter{footnote}{0}

\pagebreak
\tableofcontents
\setcounter{page}{1}


\section{Introduction}
Conformal symmetry has been extensively studied over the years because of its relevance to various aspects of theoretical physics: two-dimensional conformal theory models and statistical mechanics; four-dimensional $N=2$ and $4$ superconformal field theories; as an underlying symmetry that may be broken to Poincar\'e  symmetry in four-dimensional spacetime, and as  a tool to construct off-shell supergravity theories. It  played an important role from early on in  studies of quantum gravity \cite{Capper:1975ig}.
Here we shall be interested in the geometry of local (super) conformal theories {as represented on bundles of supertwistors over superspace}.

The first paper on $D=4, N=1$ conformal supergravity (CSG) \cite{Kaku:1978nz} used a formalism in which the entire superconformal group was gauged in a spacetime context. Although this was not a fully geometrical
set-up because supersymmetry does not act on spacetime itself, but rather on the component fields, this paper nevertheless introduced the idea that gauging conformal boosts and
scale transformations could be very useful. After Poincar\'e supergravity had been constructed in conventional (i.e. Salam-Strathdee \cite{Salam:1974jj}) superspace \cite{Wess:1977fn},\cite{Siegel:1978mj}, it was subsequently shown how scale transformations could be incorporated as super-Weyl transformations \cite{Siegel:1978mj,Howe:1978km}. On the other hand, in the completely different approach to superspace supergravity of \cite{Ogievetsky:1979ay}, super scale transformations were built in right from the start. However, it has turned out to be difficult to extend the latter approach to other cases, such as higher dimensions or higher $N$. 
The superspace geometry corresponding to all $D=4$ off-shell CSG multiplets was given in \cite{Howe:1980sy}, \cite{Howe:1981gz}
 using conventional superspace with an $SL(2,\bbC) \xz U(N)$ group in the tangent spaces together with real super-Weyl transformations. The superspace geometries corresponding to most other off-shell CSG multiplets have also been described, from the conventional point of view in $D=3$ \cite{Howe:1995zm,Kuzenko:2011xg,Gran:2012mg} and from conformal superspace in $D=3 ,4$ and $5$  {{\cite{Butter:2009cp},}}\cite{Butter:2013goa},\cite{3to5},{\cite{Kuzenko:2008wr}},\cite{Butter:2014xxa}, and in $D=6$ (the $(1,0)$ theory) \cite{Butter:2016qkx}, \cite{Butter:2017jqu}, \cite{Butter:2018wss}. { The $D=4$ $N=2$ theory \cite{Galperin:1987ek} and the $D=6$ $(1,0)$ theory were also discussed earlier in harmonic superspace in \cite{Galperin:1987ek} and \cite{Sokatchev:1988nk} respectively. In addition, some years ago, the $D=6$ $(1,0)$ theory  was formulated in in projective superspace \cite{Linch:2012zh}. }This theory as well as the $D = 6~(2, 0)$ theory were recently discussed in terms of local supertwistors in \cite{Howe:2020xrg}.
    
In the non-supersymmetric case a standard approach is to work with conventional Riemannian geometry augmented by Weyl transformations of the metric. The Riemann tensor splits in two parts, the conformal Weyl tensor, and the Schouten tensor which is a particular linear combination of the Ricci tensor and the curvature scalar. This object transforms in a connection-type way under Weyl transformations and can be used to construct a new connection known as the tractor connection \cite{Thomas, Rocky, Gover:2008sw}. This takes its values in a parabolic subalgebra of the conformal algebra and acts naturally on a vector bundle whose fibres are $\bbR^{1,n+1}$ (in the Euclidean case), thus generalising in some sense the standard conformal embedding of flat $n$-dimensional Euclidean space. Although this idea does not carry over straightforwardly to the supersymmetric case, a similar construction, which does, can be made by replacing the $(n+2)$-dimensional fibre by the relevant twistor space. This formalism is called the local twistor formalism and was introduced in $D=4$ in \cite{Penrose:1986ca}. It has been discussed in the supersymmetric case\footnote{{ Global supertwistors and the relation to conformal supersymmetry were introduced in \cite{Ferber:1977qx}}} for $N=1,2$ in $D=4$ by Merkulov, \cite{Merkulov:1986yr, Merkulov:1988zb, Merkulov:1989im}. Such a formalism depends on the dimension of spacetime because the twistor spaces also change. 

In the following we shall take a slightly different approach to that of Merkulov in that we start from a connection taking its values in the full superconformal algebra. The conformal superspace formalism alluded to previously { \cite{Butter:2009cp}, \cite{Butter:2013goa}} is a supersymmetric version of the Cartan connection formalism \cite{cartan}, which was first mentioned in the superspace context in \cite{Lott:2001st}. The formalism we advocate here can be thought of as an associated Cartan formalism in that the connection acts on a vector bundle rather than a principal one. 

In section 2 we briefly discuss bosonic conformal geometry starting from a local scale-symmetric formulation from which we recover the standard formalism; we also review the r\^ole of the Schouten tensor. We then give a brief outline of the local twistor point of view in $D=3,4,6$. A connection for the full superconformal group in the twistor representation is given and the standard formalism is recovered by  a suitable choice of gauge with respect to local conformal boost transformations. This is then generalised  to the supersymmetric case starting with a quasi-universal discussion in section 3, and followed up by details of the $D=3,4$ and $6$ cases in sections 4,5 and 6 respectively. { In section 7, the $D=5, N=1$ case, which is slightly different, is discussed.  In all cases constraints have to be imposed in order to reduce the component field content to that of the respective minimal off-shell conformal supergravity multiplets, the main one being that the even torsion two-form should be the same as in the flat case. In section 8, we describe the minimal formalism, in which one introduces only the dimension-zero torsion. This formalism applies equally well in all cases, including  $D=5$ (N=1). In section 9 we introduce a new class of supermanifolds which we call local super Grassmannians,  which could be useful in constructing alternative approaches to the subject. In section 10, we make some concluding remarks.}
\section{Conformal geometry}

From a mathematical point of view a conformal structure in $n$ dimensions can be thought of as $G$-structure with $G=CO(n)$, the orthogonal group augmented by scale transformations, see, for example \cite{Rocky, Gover:2008sw}. This group does not preserve a particular tensor, but only a metric up to a scale transformation, so that angles but not lengths are invariant. Similarly, a classical non-gravitational theory of fields which is locally scale invariant in a background gravitational field will be conformally invariant with respect to the standard conformal group of flat spacetime when the metric is taken to be flat. So conformal transformation are in this sense already present in a theory which is $CO(n)$ invariant. Nevertheless, it can sometimes be of use to make conformal boosts manifest. This can be done, for example, in the Cartan formalism which we shall describe shortly. We shall start from a Weyl perspective in which the structure group is $CO(n)$, so that, in addition to the usual curvature, there is also a scale curvature. It can easily be seen that the torsion is invariant under shifts of the scale connection, and from there, one can either use this shift symmetry to set the scale connection to zero, or introduce a new connection in order to make the curvature invariant. The first method leads back to the conventional approach, whereas the second leads to the Cartan formalism with the symmetric part of the conformal boost connection identified with the Schouten tensor, which will be defined below. 

 Let $e_a$ denote a local basis for the tangent space with dual basis forms denoted by $e^a$. So $e_a=e_a{}^m \del_m;\ e^a=dx^m e_m{}^a$, where $x^m$ denotes local coordinates and where the coordinate and preferred bases are related to each other by the vielbein $e_m{}^a$ and its inverse $e_a{}^m$. Infinitesimal local $\gc\go(n)$ transformations act on the frames by
\be
\d e_a=-\hat L_a{}^b e_b\qquad : \d e^a=e^b\hat L_b{}^a,
\ee
where 
\be
\hat L_a{}^b= L_a{}^b + \d_a{}^b S\ .
\ee
 The parameter $L_a{}^b$ denotes a local $\go(n)$ transformation preserving the flat metric $\h_{ab}$, and $S$ is a local scale transformation, although we note that $\h_{ab}$ is not invariant under the latter.

 The torsion 2-form is defined by
\be
T^a =\hat De^a=d e^a+ e^b \o_b{}^a + e^a \o_0:=D e^a + e^a \o_0\ ,
\ee
where $\o_a{}^b$ is the $\go(n)$ connection and $\o_0:=e^a \o_a$ the scale connection. In components
\be
T_{ab}{}^c=f_{ab}{}^c + 2(\o_{[a,b]}{}^c -\d_{[a}{}^c \o_{ b]})\ ,
\ee
where 
\be
f_{ab}{}^c=2 e_{[a}{}^m e_{b]}{}^n \del_m e_n{}^c\ .
\ee
It is clear that the torsion is invariant under shifts of $\o_{a}$ by a parameter $X_a$, say, provided that the $\go(n)$ connection tranforms by
\be
\o_{a,b}{}^c\rightarrow\o_{a,b}{}^c + \d_a{}^c X_b-\h_{ab} X^c\ .
\ee
It is also clear that we can use $\o_{a,b}{}^c$ to set the torsion to zero while still maintaining this symmetry, and from now on we shall take this to be the case. 

The curvature 2-form is given by $\hat R_a{}^b=R_a{}^b + \d_a{}^b R_0$ where $R_0$ is the scale curvature. Making use of the first Bianchi identity we find
\be
R_{ab,}{}^{cd}=C_{ab,}{}^{cd} + 4\d_{[a}{}^{[c} Q_{b]}{}^{d]}
\ee
where the trace-free part $C_{ab,cd}$ is the Weyl tensor and where
\be
Q_{ab}=P_{ab} + \half (R_0)_{ab}\ ,
\ee
with the symmetric part $P_{ab}$ being the Schouten tensor (also known as the rho-tensor).  It is given by
\be
P_{ab}= \frac{1}{n-2}\left(R_{ab} -\frac{1}{2(n-1)} \h_{ab} R\right)\ .
\ee
Here $R_{ab}$ is the usual symmetric Ricci tensor and R the curvature scalar. Under a finite shift of $\o_a$ accompanied by a (2.6) transformation we find
\be
\D Q_{ab}=\hat D_a X_b- X_a X_b +\half \h_{ab} X^2\ ,
\la{qtf}
\ee
where $\D$ denotes a finite change. 

At this stage one option is to use $X$ to set the scale connection to zero. This gauge will be preserved by a combined $X$ transformation and a scale transformation provided that
\be
X_a=Y_a:=S^{-1}D_a S= e_a{}^m (S^{-1} \del_m S)=e_a{}^m Y_m\ ,
\la{}
\ee
where $S$ here denotes a finite scale transformation. So at this point we have regained the conventional formalism: the local scale transformations are no longer regarded as part of the tangent space group but instead are simply rescalings of the metric without an additional scale connection. The transformation of $P_{ab}$ is given by \eq{qtf} but with $X$ replaced by $Y$, accompanied by an overall factor of $S^{-2}$. This factor then disappears in a coordinate basis so that the usual formula is recovered.  

In the conventional formalism one can define a new connection, the tractor connection, which takes its values in the Lie algebra of the conformal algebra $\gs\go (n+1,1)$, and which acts naturally to give a covariant derivative acting on vector fields in $\bbR^{n+1,1}$. 
However, the tractor formalism cannot be  adapted directly to the supersymmetric case because the superconformal groups are not simply given by  super Lorentz groups in two higher dimensions, one of which is timelike.  Instead, one should think about supertwistors because they naturally carry the fundamental representations of superconformal algebras. It is therefore more relevant to study local twistor connections, introduced in \cite{Penrose:1986ca} in the non-supersymmetric case. Then one has to consider different twistor spaces according to the dimensions of spacetime. In general we can write an element of the conformal Lie algebra, $h$, in the form
\be
h=\left(\barr{cc}
-a^\a{}_\b & b^{\a\b'}\\
c_{\a'\b} & d_{\a'}{}^{\b'}
\earr\right)
\ee
where $\a,\a'$ etc denote spinor indices which have two components for $D=3,4$ and four components for $D=6$. The diagonal elements $a,d$ are Lorentz and scale transformations while the off-diagonal ones are translations, $b$, and conformal boosts $c$. 

 The (Lie algebra valued) connection is
\be
\cA=\left(\barr{cc} -\hat\o & e\\
f&\hat\o'\earr\right)
\ee
where each entry is a one-form with the index structure given in the previous equation. The diagonal elements are connections for Lorentz and scale transformations, while the off-diagonal entries are the vielbein form $e$ and the conformal connection $f$. The transformation of $\cA$ is 
\be\label{atf}
\cA\mapsto g^{-1}\cA g+dg^{-1} g\ ,
\la{2.14}
\ee
and the curvature $\cF$ is defined by
\be
\cF=d\cA + \cA^2\ .
\ee
It transforms covariantly under $g$ without a derivative term. Its components are given by
\be
\cF=\left(\barr{cc} -\hat\cR^\a{}_\b & \cT^{\a\b'}\\
\cS_{\a'\b} & \hat\cR_{\a'}{}^{\b'}\earr\right)\ ,
\la{}
\ee
where the diagonal terms are the covariant Lorentz and scale curvatures, $\cT$ is the torsion and $\cS$ the conformal curvature. In terms of the standard torsion and curvature one has
\be
\cF=\left(\barr{cc} -\hat R^\a{}_\b-e^{\a\c'} f_{\c'\b} & T^{\a\b'}\\
Df_{\a'\b} & \hat R_{\a'}{}^{\b'}-f_{\a'\c}e^{\c\b'}\earr\right)
\la{2.17}
\ee

The objective now is to construct an element of the conformal group depending on the { scale and conformal parameters}, and a connection one-form with values in the conformal algebra which will transform in the required way provided that the transformation of the Schouten tensor is as given above { in \eq{qtf}}. The group element $g$ is given by
\be\label{gel}
g=\left(\barr{cc} S^{-\half} & 0\ ,\\
S^{-\half} C & S^{\half}\earr\right)\ ,
\ee
where the diagonal elements involve unit  matrices and $C$ is a covector, $C_{\a'\b}$. Under a conformal transformation of this form one can straightforwardly compute the changes in the components of $\cA$ to be
\begin{align}\label{219}
\hat\o &\mapsto \hat \o-eC-\half Y \ ,\w1
\hat\o'&\mapsto \hat\o'-Ce-\half Y\ ,\w1
e&\mapsto eS\ ,\w1
f&\mapsto S^{-1}(f-\hat D C+ CeC)\ ,\label{222}
\end{align}
where again the index structure on the various elements follows from the original definitions, and where $Y=S^{-1} dS$.  In order to compare with the previous discussion we need to eliminate the scale curvature and express the conformal boost parameter in terms of the scale parameter $S$. {In addition, we use the Lorentz connection to set the torsion to zero as usual.} The conformal connection $f$ is a covector-valued one-form, $f_b=e^a f_{ab}$, and we can use the anti-symmetric part of $f_{ab}$ to set $(\cR_0)_{ab}=0$. If we take the trace of the transformation of $\hat\o$, which is proportional to the transformation of $\o_0$, we can see that the conformal boost parameter can be used to set $\o_0=0$, so that $R_0=0$ as well. This gauge will be preserved if $C_a\propto Y_a$. This gives us the desired result:  the scale curvatures $\cR_0$ and $R_0$ are both zero as is the antisymmetric part of $f$ so that we can identify the remaining symmetric part $f_{(ab)}$ with the Schouten tensor $P_{ab}$.

These transformations agree with the usual ones in the conventional formalism, expressed in spinor notation and with respect to an orthonormal basis. For the connection form $\o_a{}^b$ this translates to
\be
\o_a{}^b\mapsto e^b Y_a-e_a Y^b\ .
\la{}
\ee

For $P_{ab}$ we recover the standard transformation for the tractor connection which in an orthonormal basis reads 
\be
P_{ab}\mapsto S^{-2}(D_a Y_b + Y_a Y_b-\half \h_{ab} Y^2)\ .
\la{}
\ee

Now let us return to the Weyl picture with non-zero scale connection but still with the torsion taken to be zero. If we redefine the Lorentz and scale curvature two-forms by
\begin{align}
R^{ab} &\mapsto R^{'ab}=R^{ab}+2 e^{[a}\wedge Q^{b]}\nn\w1
R_0&\mapsto R'_0=R_0-e^a\wedge Q_a\ ,
\la{}
\end{align}
where $Q^b:=e^a Q_a{}^b$, then we observe that the primed quantities are invariant under infinitesimal $X$ gauge transformations, for which 
\be
\d Q^a=\hat D X^a\ .
\la{}
\ee
We can also define a new curvature two-form
\be
R'^a:=\hat D Q^a\ .
\la{}
\ee
We can interpret $X$ as a local conformal boost parameter, $Q^a$ as the corresponding gauge field and $R'^a$ as its curvature. We have thus arrived at the conformal gauging picture starting from the Weyl perspective. {In fact, we can identify the combined primed curvatures together with the $\hat\cR$ curvatures, and $Q$ with $f$, in \eq{2.17}}. Of course, this is just the converse to deriving the conventional point of view starting from the conformal perspective as discussed, for example, in {\cite{Butter:2013goa}}
\bigskip

To conclude this outline of non-supersymmetric conformal geometry we briefly review the theory of Cartan connections of which conformal gauging is an example.
Let $H,G$ be Lie groups, $H\subset G$, with respective Lie algebras $\gh,\gg$  let $P$ be a principal $H$-bundle over a base manifold $M$. A Cartan connection on $P$ is a $\gg$-valued form $\o$ equivariant with respect to $H$, and such that $\forall X\in \gh$ $\o(X)=X$ and $\o$ gives an isomorphism from $T_p P$ to $\gg$, for any point $p\in P$.

A simple example is given by an $n$-dimensional manifold $M$ with $G=SO(n)\ltimes \bbR^n$, $H=SO(n)$. Then $\gg=\gg_{-1}\oplus\gh$, where $\gg_{-1}$ corresponds to translations and $\gh$ to rotations. The translational part of $\o$ is identified with the soldering form, \ie the vielbein, while the $\gh$-part corresponds to an $\gs\go(n)$ connection. In the conformal case $\gh=\gg_0\oplus\gg_1$ where $\gg_0=\gs\go(n)\oplus \bbR$, and $\gg_1=\bbR^n$. So $\gg_o$ corresponds to rotations and scale transformations while $\gg_1$ corresponds to conformal boosts. The grading of the Lie algebra then corresponds to the dilatational weights of the various components. The curvature of $\o$, $\cR=d\o+\o^2$, also has components corresponding to this grading, and it is straightforward to see that they correspond to the torsion, the curvature and scale curvature, and the conformal boost field strength respectively. 


\section{Superconformal geometry}

In this section we shall present a quasi-universal formalism for superconformal geometries in $D=3,4$ and $6$, working in complexified spacetime for the moment. The basic idea is to introduce a connection one-form $\cA$ on superspace
taking its values in the appropriate superconformal algebra acting on the super vector bundle whose fibres are super-twistors. These have the form
\be
Z^{\underline{\a}}=\left(\begin{array}{c}
u^\a\\
v_{\a'}\\
\l_i\\
\end{array}\right)\ .
\ee
The spinor indices on $(u,v)$ are two-component for $D=3,4$ and four-component for $D=6$. The primed spinor indices are dotted indices for $D=4$ and are the same as the unprimed
ones in all other cases. The odd part of a super-twistor is $\l_i$, where $i=1,\ldots M$, with $M=N$, the number of supersymmetries for $D=3,4$ and $M=2N$ for $D=6$. 

The connection is given by
\be
\cal{A}=\left(\begin{array}{c c |c}
-\hat\O^\a{}_{\b}\ &E^{\a\b'} & E^{\a j}\\
F_{\a'\b} & \hat{\O}_{\a'}{}^{\b'}&F_{\a'}{}^j\\
\hline
\tilde F_{i\b}&\tilde E_i{}^{\b'}& \O_i{}^j
\end{array}\right)\ .
\ee
 For $D=3$ the pair ($\a\b)$ on $(E,F)$ are symmetric whereas in $D=6$ they are antisymmetric. The connection components are as follows: ($E^{\a\b'},E^{\a j}$) correspond to translations and $Q$-supersymmetry, $(F_{\a'\b},F_{\a'}{}^j)$ to conformal and $S$-supersymmetry, while the diagonal $\O$s
correspond to Lorentz symmetry above the line and internal symmetry below the line with the hats indicating that the scale connection is also included. $(\tilde{E},\tilde{F})$ on the third line are appropriate transpositions of $(E,F)$. We will occasionally refer to the internal connection as $\O_I$. 

The curvature two-form is given by $\cF=d\cA+\cA^2$; its components in matrix form are:
\be
\cal{F}=\left(\begin{array}{c c |c}
-\hat \cR^\a{}_{\b}\ &\cT^{\a\b'} & \cT^{\a j}\\
\cS_{\a'\b} & \hat{\cR}_{\a'}{}^{\b'}&\cS_{\a'}{}^j\\
\hline
\tilde\cS_{i\b}&\tilde{\cT}_i{}^{\b'}& \cR_i{}^j
\end{array}\right)\ ,
\ee
where

\begin{align}\label{aha}
\cT^{\a\b'}&=\hat T^{\a\b'}+E^{\a k} \tilde{E}_k{}^{\b'}\nn\w1
\cT^{\a j}&=\hat  T^{\a j} + E^{\a \c'}F_{\c'}{}^{j} \nn\w1
\cS_{\a'\b}&=\hat D F_{\a'\b} +F_{\a'}{}^k \tilde{F}_{k\b}\nn\w1
\cS_{\a'}{}^j&=\hat D F_{\a'}{}^j +F_{\a'\c}E^{\c j}\nn\w1
\hat{\cR}^\a{}_\b&=\hat R^\a{}_\b + E^{\a\c'} F_{\c'\b} +E^{\a k} \tilde{F}_{k\b}\nn\w1
\hat{\cR}_{\a'}{}^{\b'}&=\hat R_{\a'}{}^{\b'} + F_{\a'\c} E^{\c\b'} +F_{\a'}{}^k \tilde{E}_{k}{}^{\b'}\nn\w1
\cR_i{}^j&= R_i{}^j +\tilde{F}_{i\c} {E}^{\c j} +\tilde{E}_i^{\c'}F_{\c'}{}^j \ .
\end{align}

Here $\hat D$ denotes the exterior covariant derivative for the Lorentz, scale and internal parts of the algebra. The first terms on the right of the first two equations are the standard even and odd superspace torsion tensors constructed in the usual way from the $\hat\O$ connections,  $\hat T^{\a\b}=\hat D E^{\a\b}$, $\hat T^{\a j}=\hat D E^{\a j}$, while the non-calligraphic curvature forms on the right in the last three lines are the standard curvature tensors constructed in a similar fashion.

In $D=3,6$ the one-forms $(E^{\a\b'},E^{\a j})$ can be identified, after converting pairs of  spinor indices to vector indices and, if necessary, rescaling, with the basis one-forms of conventional superspace, $E^A=(E^a,E^{\a i})$, while in $D=4$, $\tilde E_i{}^{\a'}$, which becomes the complex conjugate of $E^{\a i}$ in real superspace, is also required in order to complete the basis forms.  The forms $E^A$ are associated with super-translations, and play the role of soldering forms in this context. This means that the translational part of the algebra can be subsumed into 
super-diffeomorphisms.

 In a similar fashion, we can identify the pair $(F_{\a'\b}, {F}_{\a'}{}^{ j})$ with a super-covector-valued one-form $F_B=E^A F_{AB}$. In $D=4$ the tilded odd forms are essentially the complex 
conjugates of the untilded ones (in real spacetime) and similar identifications can be made in other dimensions.

{ We remark in passing that, although there is also a conformal supergravity theory in $D=5$ for $N=1$ {\cite{Kugo:2002vc},\cite{Bergshoeff:2002qk}}, which can be described in conventional superspace {\cite{Kuzenko:2008wr} and in conformal superspace {\cite{Butter:2014xxa}}}, it does not admit a straightforward description in the supertwistor formalism. In this case the spinor indices are four-component, with no distinction between primed and unprimed, vectors can be represented by skew-symmetric symplectic-traceless spinors, \eg $E^{\a\b}=-E^{\b\a},\,\h_{\a\b} E^{\a\b}=0$, where $\h_{\a\b}$ is the symplectic matrix (charge conjugation matrix), and where the spacetime part of the curvature is $\c^a$-traceless, $\hat\O^\a{}_\b (\c^a)^\b{}_\a=0$. However, the bilinear fermion terms in the bosonic curvatures in \eq{aha} do not preserve these contraints.}  {Nevertheless, we shall see in section 7, that the formalism can be amended to take the $D=5$ case into account.}

The Bianchi identity is 
\be
\cD\cF:= d\cF + [\cF,\cA]=0\ .
\la{4.13.1}
\ee
Written out in components this is, for the torsions,
\begin{align}
\hat D \cT^{\a\b'}-E^{\a\c'} \hat\cR_{\c'}{}^{\b'}- \hat\cR^\a{}_{\c}E^{\c\b'}      -i E^{\a k} \tilde{\cT}_k{}^{\b'}-i\cT^{\a k} \tilde{E}_k{}^{\b '}&=0\nn\w1
\hat D\cT^{\a j}+\hat \cR^\a{}_\c E^{\c j}-E^{\a k}\cR_k{}^j+i\cT^{\a\c'} F_{\c'}{}^j-iE^{\a\c'} \cS_{\c'}{}^j&=0\ ,
\label{4.13.1}
\end{align}
for the Lorentz, scale and internal symmetry curvatures,
\begin{align}
D \hat \cR^\a{}_{\b} -\cT^{\a\c'} F_{\c'\b} -\cT^{\a k} \tilde{F}_{k \b }+ E^{\a\c'} \cS_{\c'\b} + E^{\a k} \tilde{\cS}_{k \b}&=0\nn\w1
 D\cR_{ij}-2\cT^\c_{(i} F_{|\c|j)} +2 E^\c_{(i} \cS_{|\c| j)}&=0\ ,
 \la{4.13.3}
 \end{align}
 and for the superconformal curvatures,
 \begin{align}
 \hat D\cS_{\a'\b} +\hat\cR_{\a'}{}^{\c'} F_{\c'\b} + F_{\a'\c} \cR^{\c}{}_{\b} + i \cS_{\a'}{}^k \tilde{F}_{k\b}+i F_{\a'}{}^k \tilde{S}_{k\b}&=0\nn\w1
 \hat D\cS_{\a'}{}^j +\hat\cR_{\a'}{}^{\c'} F_{\c'}{}^j-F_{\a'}{}^k \cR_\k{}^j +i\cS_{\a'\c} E^{\c j}-iF_{\a'\c} T^{\c j}&=0\ .
 \la{4.13.4}
 \end{align}

 We now consider the supersymmetric counterpart of the group element \eq{gel} given by
\be
g=\left(\begin{array}{c c |c}
S^{-\half}& 0& 0\\
S^{-\half} C& S^{\half}& \C\\
\hline
S^{-\half} \D& 0& 1
\end{array}\right)
\la{4.14}
\ee 
with inverse given by
\be
g^{-1}=\left(\begin{array}{c c |c}
S^{\half}& 0& 0\\
S^{-\half} \tilde C& S^{-\half}&-S^{-\half} \C\\
\hline
- \D& 0& 1
\end{array}\right)\ ,
\la{4.15}
\ee
where 
\be
C+\tilde C-\C\D=0 \ .
\la{}
\ee 
For the moment we can think of $g$ as an element of the complex supergroup $SL(M_0|M_1)$, where $(M_0|M_1)$ denote the even and odd dimensions of supertwistor space. In the real cases we will find that $\D$ is related to $\C$ and that $C$ will have symmetry properties depending on the  case in hand. The transformation of the connection $\cA$ is given by \eq{atf}. For the superspace basis forms we find
\begin{align}
E^{\a\b'}&\mapsto S E^{\a\b'}\w1\nn
E^{\a j}&\mapsto S^{\half}(E^{\a j} + E^{\a\c'} \C_{\c'}{}^j)\w1\nn
\tilde E_i{}^{\b'}&\mapsto S^{\half}(\tilde E_i{}^{\b'}-\D_{i\c} E^{\c \b'})\ ,
\la{}
\end{align}
while for the connections we find
\begin{align}
\label{contran}
\hat\O^\a{}_\b&\mapsto \hat\O^\a{}_\b-E^{\a\c'} C_{\c'\b} -E^{\a k} \D_{k\b}-\half \d^\a_\b Y\w1\nn
\hat\O_{\a'}{}^{\b'}&\mapsto \hat\O_{\a'}{}^{\b' }+\tilde C_{\a' \c} E^{\c \b'}-\C_{\a'}{}^k \tilde E_k{}^{\b'} -\half \d^{\a'}_{\b'} Y\w1\nn
\O_i{}^j &\mapsto \O_i{}^j -\D_{i \c} E^{\c j} +\tilde E_i{}^{\c'} \C_{\c'}{}^j -\D_{i\c} E^{\c\c'} \C_{\c'}{}^j ,
\end{align}
For the $S$-{ super}symmetry connections we have
\begin{align}
F_{\a'}{}^j&\mapsto S^{-\half}\left(F_{\a'}{}^j -\hat D\C_{\a'}{}^j+\tilde C_{\a'\b}(E^{\b j} +E^{\a\c'}\C_{\c'}{}^j)-\C_{\a'}{}^k \tilde E_k{}^{\b'}\C_{\b'}{}^j\right)\w1\nn
 \tilde F_{i\b}&\mapsto S^{-\half}\left(\tilde F_{i\b}-\hat D\D_{i\b}+(\tilde E_i{}^{\c'}-\D_{i\c}E^{\c\c'})C_{\c'\b}-\D_{i\c}E^{\c j}\D_{j\b}\right)\ ,
\la{}
\end{align}
The conformal connection, $F_{\a'\b}$, transforms as
\begin{align}
F_{\a'\b}&\mapsto S^{-1} (F_{\a'\b}-\hat D C_{\a'\b}+\C_{\a'}{}^k \hat D\D_{k\b} +F_{\a'}{}^k\D_{k\b}-\C_{\a'}{}^k\tilde F_{k\b}\w1\nn
&+\tilde C_{\a'\c} E^{\c\d'} C_{\d'\b} +\tilde C_{\a'\c} E^{\c k} \D_{k\b}-\C_{\a'}{}^k \tilde E_k{}^{\d'} C_{\d'\b})\ .
\la{}
\end{align}

The transformations of the field strengths can also be found straightforwardly. For the torsions we have
\begin{align}\label{tortf}
\cT^{\a\b'}&\mapsto S\cT^{\a\b'}\w1\nn
\cT^{\a j}&\mapsto S^{\half}(\cT^{\a j} + \cT^{\a\c'} \C_{\c'}{}^j)\w1\nn
\tilde \cT_i{}^{\b'}&\mapsto S^{\half}(\tilde \cT_i{}^{\b'}-\D_{i\c} \cT^{\c \b'})\ ,
\la{}
\end{align}
while for the Lorentz, scale and internal curvatures we have:
\begin{align}
\hat\cR^\a{}_\b&\mapsto \hat\cR^\a{}_\b-\cT^{\a\c'} C_{\c'\b} -\cT^{\a k} \D_{k\b}\w1\nn
\hat\cR_{\a'}{}^{\b'}&\mapsto \hat\cR_{\a'}{}^{\b' }+\tilde C_{\a' \c} \cT^{\c \b'}-\C_{\a'}{}^k \tilde \cT_k{}^{b'} \w1\nn
\cR_i{}^j &\mapsto \cR_i{}^j -\D_{i \c} \cT^{\c j} +\tilde \cT_i{}^{\c'} \C_{\c'}{}^j -\D_{i\c} \cT^{\c\c'} \C_{\c'}{}^j ,
\end{align}

The scale curvatures for $D=3,6$ are given by the trace of $\hat\cR^{\a}{}_{\b}$ given in \eq{aha}, while for $D=4$ we have to take the sum of the traces of $\hat\cR^\a{}_\b$ and its complex conjugate; in all cases, we can write
\be
(\cR_0)_{AB}=(d\O_0)_{AB} +k F_{[AB]}\ , 
\ee
where $k$ is a constant depending on the dimension of spacetime, and where $F_{[AB]}$ is the graded anti-symmetric part of $F_{AB}$, the latter having no symmetry. Since $F_A$ is a connection one-form we are free to add a tensorial part to it and thereby set the scale curvature $\cR_0=0$, after which $(d\O_0)_{AB}=-F_{[AB]}$. As in the bosonic case $\O_0$ transforms by a shift under conformal and $S$-supersymmetry transformations, as can be seen from { the trace of the first line in \eq{contran}}, and can therefore also be set to zero. We are then left with the symmetric part of $F_{AB}$ which we can be identified as the super Schouten tensor. There is then a residual local super-Weyl invariance which {we will }present in more detail below for each case.

To put more flesh on this general outline we shall now go through the various cases in turn.

\section{$D=3$}

The superconformal group in $D=3$ is $SpO(2|N)$. This is the same as the orthosymplectic group, but with the symplectic factor written first to indicate that it refers to the spacetime part. It acts on the
supertwistor space $\bbC^{4|N}$ (in the complex case) and consists of $(4|N)\xz (4|N)$ matrices $g$ which preserve the symplectic-orthogonal form
\be
J=\left(\begin{array}{c c |c}
0&1&0\\
-1 & 0& 0\\
\hline
0&0& 1_N
\end{array}\right)\ .
\ee
The invariance condition is
\be
gJ g^{st}=J
\ee
where the $st$ superscript denotes the super-transpose of the matrix $g$. This is the same as the ordinary transpose except for the odd component in the lower left corner which has an additional minus sign.
For a Lie superalgebra element $h$ we have, correspondingly,
\be
hJ=-J h^{st}
\ee
The reality condition needed to restrict to real spacetime (and superspace) is
\be
gK g^*=K
\ee
where $K$ has a similar structure to J but with the minus sign on the second row replaced by a plus sign. The form of a real superalgebra element $h$ is therefore
\be
h=\left(\begin{array}{c c |c}
a&ib&\c\\
ic & d&\d \\
\hline
\d^t&-\c^t& e
\end{array}\right)\ ,
\ee
where $d=-a^t$, $b$ and $c$ are symmetric, $e$ is anti-symmetric, $\c$ is real and $\d$ is imaginary. Since the connection is a $\gg$-valued one-form it can be written as
\be
\cal{A}=\left(\begin{array}{c c |c}
-\hat\O^\a{}_{\b}\ &iE^{\a\b} & E^{\a j}\\
iF_{\a\b} & \hat{\O}_{\a}{}^{\b}&F_{\a}{}^j\\
\hline
\tilde{F}_{i\b}&\tilde{E}_i{}^{\b}& \O_i{}^j
\end{array}\right)\ .
\ee
Here $E^{\a\b}$ and $F_{\a\b}$ are real and symmetric, $E^{\a j}$ is real and $F_\a{}^j$ is imaginary (for later convenience). { On the bottom row $\tilde{E}_i{}^\b=-E^\b{}_i$, while $\tilde{F}_i{}^\b=F^\b{}_i$, where the $\go(N)$ indices {are} raised or lowered by the flat Euclidean metric.} The Lorentz and scale connections are
\be
\hat\O^\a{}_\b=\O^\a{}_\b +\half \d^\a_\b \O_0;\qquad  \hat\O_\a{}^\b=\O_\a{}^\b +\half \d^\b_\a \O_0\ ,
\ee
with $\O^\a{}_\b=\O_\b{}^\a$ as the trace-free Lorentz connection and $\O_0$ the scale connection. As usual two-component spinor indices are raised and lowered with the epsilon tensor, so $\O_{\a\b}$ is symmetric.

It is straightforward to compute the components of the curvature two-form, but for the moment we shall focus on the scale curvature. It is given by
\be
\cR_0=R_0+E^{\a\b} F_{\a\b} -E^{\a i} F_{\a i}\ ,
\ee
where $R_0=d\O_0$. The last two terms can be rewritten as
\be
E^{\a\b} F_{\a\b} -E^{\a i} F_{\a i}=- E^a F_a -E^{\a i} F_{\a i} := -E^A F_A=-E^B E^A F_{AB}\ ,
\ee
Note that $F_{AB}$ is not necessarily graded-antisymmetric, although it is when contracted with two sets of basis forms. (We have suppressed the wedge symbol in the two-forms above). Here we have identified $E^{\a i}$ as the standard odd basis forms of superspace and set $E^{\a\b}=-E^a (\c_a)^{\a\b}$, where $E^a$ are the standard even basis forms. (Note that this involves a rescaling since
there would normally be a factor of $\half$ when going from bi-spinors to vectors.) We therefore have
\be
(\cR_0)_{AB}=(R_0)_{AB}-2 F_{[AB]}\ .
\ee
As discussed briefly above, we can use the freedom to adjust a connection by a tensorial addition to set $(\cR_0)_{AB}=0$, after which $(R_0)_{AB}=2F_{[AB]}$. As we shall see shortly below, superconformal gauge transformations (i.e. conformal boosts and S-supersymmetry)  can be used to set $\O_0=0$, after which $F_{AB}$ will become graded symmetric. We can then identify $F_{AB}$ as the super-Schouten tensor.

We shall now exhibit a finite superconformal transformation which will allow us to set $\O_0=0$ and to identify residual superconformal transformations in terms of scale transformations, or what one could
call (finite) super Weyl transformations in this context. It is given by
\be
g=\left(\begin{array}{c c |c}
S^{-\half}\ & 0& 0\\
C_0 & S^{\half}&\C\\
\hline
S^{-\half} \C^t&0& 1
\end{array}\right)\ .
\ee
where
\be
C_0=-iS^{-\half}(C +\frac{i}{2} \C\C^t)\ .
\ee
The parameters $(S,C,\C)$ are those for scale transformations, conformal boosts and special supersymmetry respectively. $C$ is symmetric on its spinor indices (i.e. it is a vector) and real, while $\C$ is taken to be imaginary. It is straightforward to check that this is indeed an element of the superconformal group; one can find the effect of such a transformation on the connection by using the standard formula
\be
\cA\rightarrow dg^{-1} g + g^{-1}\cA g \ .
\ee
For the moment we shall focus on the transformation of the scale connection $\O_0$. It is given by
\bea
\O_0\rightarrow &\O_0& -S^{-1}dS -E^{\a\b} C_{\a\b} +E^{\a i} \C_{\a i}\nn\w1
=&\O_0&  - Y+ E^a C_a + E^{\a i} \C_{\a i}\ .
\eea
where the one-form $Y:=S^{-1}dS$.  We learn two things from this equation: first, the parameters $C,\C$ can be used to set the even and odd components of the one-form $\O_0$ to zero, and second, we can determine the residual symmetry transformations in terms of $Y$. In other words, having set $\O_0=0$ we have
\be\label{3dY}
C_a=Y_a=S^{-1} D_a S; \qquad \C_{\a i}=\U_{\a i}=S^{-1} D_{\a i} S
\ee
where the derivatives are now the standard superspace covariant derivatives, and where we have denoted the odd component of $Y$ by $\U$ for later use.
The finite transformations of the basis forms are:
\begin{align}
E^a&\rightarrow SE^a \w1\nn
E^{\a i}&\rightarrow S^{\half}(E^{\a i} -i E^a(\c_a)^{\a\b}\U_{\b}{}^i)
\end{align}
The finite changes of the Lorentz and $\go(N)$ connections are:
\begin{align}\label{sw3}
\D\O^{\a\b}&=E^a(\c_a)^{(\a|\c|} (Y_\c{}^{\b)}+\frac{i}{2}(\U^2)_\c{}^{\b)}) +E^{(\a i}\U_i{}^{\b)}\w1\nn
\D\O^{ij}&= 2E^{\a[i} \U_\a{}^{j]}+i E^a\U_{\a}{}^i(\c_a)^{\a\b}\U_{\b}{}^j
\end{align}
The transformations of the superconformal connections are
\begin{align}
F_{\a}{}^j&\mapsto S^{-\half}\left(F_{\a}{}^j - D\C_{\a}{}^j+ (i C_{\a\b}+ \C_{(\a}{}^k \C_{\b) k} ) (E^{\b j} +E^{\a\c}\C_{\c}{}^j)-\C_{\a}{}^k E^{\b}{}_k{} \C_{\b}{}^j\right)
 \la{}
\end{align}
and
\begin{align}
F_{\a\b}&\mapsto S^{-1}\left(F_{\a\b} + i D C_{\a\b}-D\C_{(\a}{}^k \C_{\b)k} +2F_{(\a}{}^k\C_{k\b)}\right.\w1\nn
&\left. + C_{\a\c} E^{\c\d} C_{\d\b} + 2iC_{(\a\c} E^{\c k} \C_{k\b)}\right.\w1\nn
&\left. +iC_{(\a\c} E^{\c \d} (\C^2)_{\d\b)} +\C^2_{(\a\c} E^{\c k}\C_{\b) k} + \quart \C^2_{\a\c} E^{\c\d} \C^2_{\d\b}\right)\ .
\la{}
\end{align}
Here $\C^2_{\a\b}:=\C_{\a}{}^k\C_{\b k}$, is antisymmetric on $\a\,\b$, while the symmetrisations are on $\a,\b$ only.

The formalism given above applies quite generally regardless of whether any constraints have been imposed or not. The basic constraint we shall choose for $D=3$ (and in fact in all cases)  is $\cT_0=0$, which is clearly superconformally invariant, from \eq{tortf}. We can now use conventional constraints, including some for the conformal and superconformal potentials, as well as the Bianchi identities, to show that we can always choose $\cT_1=\cR=0$, remembering that we always choose the scale curvature to vanish. Thus in $D=3$ the only covariant field strengths that are non-zero are the internal curvature $\cR_I$ and the superconformal curvature $\cS_A=(\cS_a,\cS_{\a j})$. From the Bianchi identity \eq{4.13.1} we can then see that 
\begin{align}
\cR_{\a i\b j,kl}&=\ve_{\a\b} W_{ijkl}\ ,\w1\nn
\cR_{a\b j,kl}&= (\c_a \l)_{\b jkl}\ ,\w1\nn
\cR_{ab,kl}&=F_{ab kl}\ ,
\la{}
\end{align}
for $N\geq 2$, where each field is totally antisymmetric on its $SO(N)$ indices.
In fact, for $N\geq2$, the leading component in the Weyl multiplet, \ie the conformal supergravity field strength multiplet, is the leading non-zero component in $\cR_I$.

For $N=1$ $\cR_I$ is identically zero, and only $\cS_A$ survives. From the Bianchi identities it is then easy to show that the only non-zero components are $\cS_{ab,\c}$ and $\cS_{ab,c}$. The former is equivalent to a gamma-traceless vector-spinor, while the latter is equivalent to a symmetric traceless tensor. These are the Cottino and Cotton tensors respectively. It is then straightforward to see that the component fields in the Weyl multiplet can be arranged diagrammatically as follows\footnote{{This multiplet was discussed in the context of the supermembrane in \cite{Howe:2004ib}.} }:
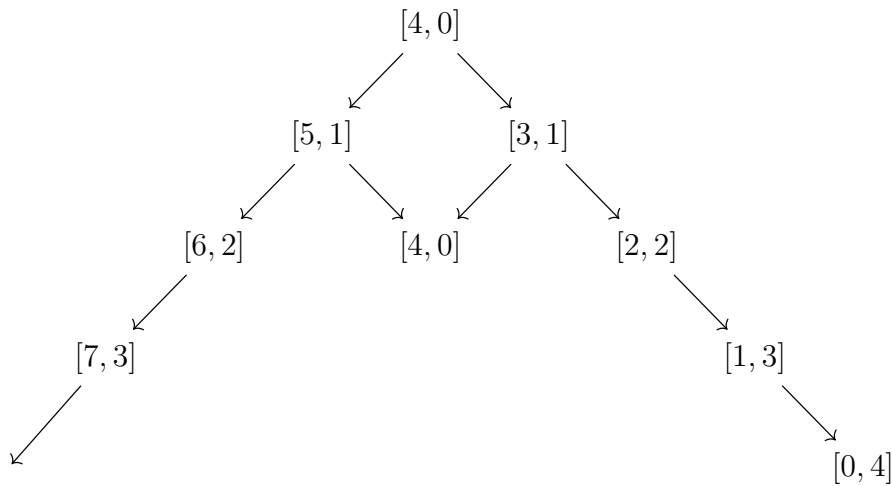
\begin{center}\begin{tikzcd}[column sep=tiny]
&&&& {[4,0]} \arrow[dl] \arrow[dr] & \\
&&&{[5,1]}\arrow[dl] \arrow[dr]& &  {[3,1]}\arrow[dl]\arrow[dr]\\
&&{[6,2]}\arrow[dl]&& {[4,0]}&&{[2,2]}\arrow[dr]\\
&{[7,3]}\arrow[dl] &&&&&&{[1,3]}\arrow[dr]\\
{~~~~}&&&&&&&&{[0,4]}
\end{tikzcd}\end{center}

\noindent Here, each $[p,q]$  entry denotes a field with $p$ antisymmetrised internal indices and $q$ symmetrised spinor indices. For $N\leq 4$ clearly only the right part of the diagram survives, with the last two components being the Cottino and Cotton tensors respectively. For $N=6$ we note that there is an extra $U(1)$ gauge field not included in the superconformal group. This field plays an important role in the BLG formalism for multiple membranes \cite{Bagger:2007jr,Gustavsson:2007vu}, and was discussed in the superspace context in \cite{Gran:2012mg}. For $N=6$, therefore, we have an additional closed two-form field strength $G$ with components
\begin{align}
G_{\a i \b j}&= \ve_{\a\b} W_{ij}\w1\nn
G_{a\b j}&=(\c_a \l)_{\b j}\ ,
\la{}
\end{align}
as well as $G_{ab}$, where $W_{ij}$ is the $SO(6)$ dual of $W_{ijkl}$ and $\l_i$ is the $SO(6)$ dual of the 5-index fermion on the left of the diagram above. For $N=8$ an extra constraint is required in order to avoid having two gravitons; this is achieved by imposing a self-duality constraint on $W_{ijkl}$, and this in turn implies that the field content of the $N=7$ and $N=8$ Weyl multiplets are the same, so that the left-hand diagonal line can be terminated at $N=6$.

To conclude this section we translate the above results into conventional superspace. The main consequence is that the components of the superconformal potentials now appear explicitly in the torsions and curvatures. These potentials are graded symmetric and make up the components of the super-Schouten tensor as we remarked earlier. Making use of equation \eq{aha} which relates the conformally covariant tensors on the left to the standard superspace ones on the right we find, for the even torsion two-form
\be
T^a=-\frac{i}{2} E^{\a i} (\c^a)_{\a\b} E^{\b}_i\ ,
\ee
or, in components,
\begin{align}
T_{\a i\b j}{}^c&=-i\d_{ij} (\c^a)_{\a\b}\ ,\nn\w1
T_{\a i b}{}^c&=T_{ab}{}^c=0\ ,
\end{align}
which we recognise as the usual expressions {\cite{Howe:1995zm}}. For the odd torsion  we find
\be
T^{\a i}=D E^{\a i}=(\c_a)^{\a \c} E^a(E^{\b j}F_{\b j,\c}{}^i + E^b F_{b\c}{}^i)\ .
\ee
This implies for the components,
\begin{align}
T_{\a i,\b j}{}^{\c k}&=0\ , \nn\w1
T_{a\b j}{}^{\c k}&=-(\c_a)^{\c\d} F_{\b j,\d}{}^k\ ,\nn\w1
T_{ab}{}^{\c k}&=-2\c_{[a}^{\c\d} F_{b] \d}{}^k ,
\end{align}
where we have written the right-hand sides in terms of the super Schouten tensor. For the dimension-one component, since $F_{\a i,\b j}$ is antisymmetric under the interchange of pairs of indices, we have
$F_{\a i,\b j}=\ve_{\a\b} K_{ij} + (\c^a)_{\a\b} L_{a ij}$, where $K$ is symmetric on the internal indices and $L$ antisymmetric,  so that
\be\label{dim1b}
T_{a\b j}{}^{\c k}= (\c_a)_\b{}^\c K_j{}^k +(\c^b\c_a)_\b{}^\c L_{b j}{}^k\ .
\ee
This differs slightly from the expression {for the dimension one torsion} given in {\cite{Howe:1995zm,Gran:2012mg}}, but can be brought into agreement by a further redefinition of the dimension-one $\go(N)$ connection. (On the other hand {\eq{dim1b}} does agree with the form given in {\cite{Kuzenko:2011xg}}.)
The dimension three-halves torsion is the gravitino field strength, which in three dimensions can be dualised to a vector-spinor $\Psi_{a \c k}$. It is given in terms of the super Schouten tensor by
\be
T_{ab}{}^{\c k} =-2(\c_{[a}){}^{\c\d} F_{b]\d}{}^k\ ,
\ee
where
\be
\Psi_a{}^{\a i}:=\half \ve_a{}^{bc}\Psi_{bc}{}^{\a i}=\ve_a{}^{bc} (\c_b)^{\a\b} F_{c \b}{}^i\ .
\ee

The components of the standard superspace curvature tensors can also be easily computed from equation (3.4), using the fact that the covariant Lorentz and scale curvatures are zero.

\section{$D=4$}

The superconformal groups for $D=4$ $N$-extended supersymmetry are $SU(2,2|N)$, for $N=1,2,3$ and $PSU(2,2|4)$ for $N=4$. Elements of this group are matrices $g$ with unit superdeterminant which obey
\be
g^* J g=J
\ee
where 
\be
J=\left(\begin{array}{c c |c}
0&1_2&0\\
1_2 & 0& 0\\
\hline
0&0& 1_N
\end{array}\right)\ ,
\ee
and where the numerical subscripts denote the dimenions of the unit matrices. For an element $h$ of the corresponding Lie superalgebra $\gs\gu(2,2|N)$ we have
\be
h^* J=-Jh
\ee
Such an element can be written
\be
h=\left(\begin{array}{c c |c}
-\hat a&-ib&-i\ve\\
-ic & \hat a^*&\vf \\
\hline
-\vf^*&-i\ve^*& e
\end{array}\right)\ ,
\ee
where $b$ and $c$ are hermitian, $e$ is anti-hermitian, and the factors of $i$ are put in for convenience (we follow the conventions of \cite{Howe:1995md} here). The supertrace condition then implies that
\be
-a_0+\bar a_0=e_k{}^k\ ,
\ee
where
\be
\hat a^\a{}_\b=a^\a{}_\b+\half \d^\a_\b a_0\ .
\ee


The connection in $D=4$ can be written
\be
\cal{A}=\left(\begin{array}{c c |c}
-\hat\O^\a{}_{\b}\ &-iE^{\a\bdt} & -i E^{\a j}\\
-iF_{\adt\b} & \hat{\bar{\O}}_{\adt}{}^{\bdt}&\bar{F}_{\adt}{}^j\\
\hline

-F^t_{i \b}&-i E^*_i{}^{\bdt}& {\O}_i{}^j
\end{array}\right)\ .
\ee
and the curvature is
\be
\cal{F}=\left(\begin{array}{c c |c}
-\hat \cR^\a{}_{\b}\ &-i\cT^{\a\bdt} & -i\cT^{\a j}\\
-i \cS_{\adt\b} & \hat{\bar\cR}_{\adt}{}^{\bdt}& \bar{\cS}_{\adt}{}^j\\
\hline
-(\cS^t)_{i\b}&-i({\cT^*})_i{}^{\bdt}& \cR_i{}^j
\end{array}\right)\ ,
\ee
where the expressions for the various components are given by (3.4) with appropriate factors of $i$.
We set
\be
\hat\O^\a{}_\b=\O^\a{}_{\b} +\half\d^{\a}_{\b} (\O_0+i\O_1)\ ,
\la{}
\ee
where $\O_0$ and $\O_1$ are the scale and $U(1)$ connections respectively and where $\O^\a{}_\b$ is trace-free.

A group element depending on scale and superconformal transformations is easily constructed in a similar fashion to $D=3$. It is given by
\be
g=\left(\begin{array}{c c |c}
S^{-\half}\ & 0& 0\\
C_0 & S^{\half}&\bar{\C}\\
\hline
-S^{-\half} \C^t&0& 1
\end{array}\right)\ .
\ee
where
\be
C_0=i S^{-\half}(C+\frac{i}{2}\bar\C\C^t)
\ee
with $C$ being hermitian. The inverse is given by
\be
g^{-1}=\left(\begin{array}{c c |c}
S^{\half}\ & 0& 0\\
\hat C_0 & S^{-\half}&-S^{-\half}\bar\C\\
\hline
\C^t&0& 1
\end{array}\right)\ .
\ee
where
\be
\hat C_0=-i S^{-\half}(C-\frac{i}{2}\bar\C\C^t)\ .
\ee

As in $D=3$ we can use the antisymmetric part of the superconformal connection together with a superconformal transformation to set $\cR_0=R_0=0$ and leave residual superconformal transformations in terms of the scale parameter $S$. The transformation of the scale connection $\O_0$ under $g$ is given by
\be
\D\O_0=-Y+E^{\a\adt} C_{\adt\a} -\frac{i}{2} E^{\a k} \C_{\a k}-\frac{i}{2}\bar{E}^{\adt}_k \bar\C_{\adt}{}^k\ .
\la{}
\ee
The right-hand side can be rewritten as $-Y+C$, where $C=E^A C_A=E^a C_a + E^{\a i} \C_{\a i} - 
\bar{E}^{\adt}_i \C_{\adt}{}^i$ (after rescaling $C_{\a\adt}\rightarrow \half C_{\a\adt}$ and $\C_{\a i}\rightarrow 2i \C_{\a i}$). We can then use $C$ to set $\O_0=0$, leaving residual superconformal transformations with $Y=S^{-1} d S=C$.

The scale curvature $\cR_0$ can be written as
\be
\cR_0=R_0-E^A f_A\ ,
\ee
where $R_0=d \O_0$ and where
\be
f_a:=2F_a; \ \ \  f_{\a i}:=-\frac{i}{2} F_{\a i};\  \ \ \bar{f}_{\adt}{}^i:= \frac{i}{2} \bar{F}_{\adt}{}^i\ .
\la{}
\ee
We therefore have
\be 
(\cR_0)_{AB}=(R_0)_{AB}-2 f_{[AB]}\ .
\la{}
\ee
Finally,  we can use $C_A$ to set $(\O_0)_A=0$ and the graded antisymmetric part of the superconformal connection $f_{[AB]}$ to set $\cR_0=0$. After this, we have residual superconformal transformations determined by $S$, as above, while the remaining part of $f_{AB}$ is graded symmetric and can be identified as the super-Schouten tensor for $D=4$.

The super-Weyl transformations of the basis forms and the connections are given by:
\begin{align}\label{sw4}
E^a&\mapsto  S E^a\ ,\nn\w1
E^{\a i}&\mapsto S^{\half} E^{\a i}\ ,\nn\w1
\O^\a{}_\b&\mapsto \O^\a{}_\b -\half E^{\a\cdt} (C_{\cdt\b}+4i\bar\U_{\cdt}{}^k\U_{\b k})+2E^{\a k}\U_{\b k}\ , \nn\w1
\O_i{}^j&\mapsto\O_i{}^j -2(E^{\c j} \U_{\c i}+\bar{E}^{\cdt}_i\bar\U_{\cdt}^j+2i\U_{\c i} E^{\c\cdt} \bar\U_{\cdt}{}^j)\ ,\nn\w1
\end{align}
where it is understood that the trace over $\a$ and $\b$ is to be projected out in the third line. The components of the super-Schouten tensor at dimension one are $f_{\a i,\b j}$, $f_{\a i,\bdt}^{\ \ \ j}$ plus complex conjugates. Graded symmetry then implies that
\be
f_{\a i,\b j}=\ve_{\a\b} f_{(ij)} + f_{(\a\b)[ij]}\ ,
\ee
while $f_{\a i,\bdt}^{\ \ \ j}$ is skew-hermitian. At dimension three-halves there is just a complex vector-spinor $f_{a \b j}$ which can be identified with that part of the gravitino field-strength tensor which is not part of the Weyl supermultiplet.

The basic constraint for $D=4$ is $\cT^a=0$, as for the other cases. Using the Bianchi identities and choices for the connections, including the superconformal ones, one finds that the non-zero components of the covariant torsion $\cT^{\c k}$ are 
\begin{align}\label{tor}
\cT_{\adt\bdt}^{i\, j\,\c k}&=\ve_{\adt\bdt}\chi^{\c ijk}\ ,\nn\w1
\cT_{\a\adt,\b,\cdt}^{ \ \ \ \,j\, k}&=i\bar M_{\a\b}^{jk}\ ,\nn\w1
\cT_{\a\adt,\b,\c k}^{\ \ \ j}&=-\frac{i}{12} (N-3) \d_k^j \bar\chi_{\adt}^{lmn}\chi_{\b lmn}\ ,\nn\w1
\cT_{\a\adt,\b\bdt,\cdt}^{\ \ \ k}&=\ve_{\a\b}\bar\Psi_{\adt\bdt\cdt}^k\ .
\end{align}
Here, vector indices have been converted to pairs of spinor indices using the sigma-matrices in the usual way:
\be
\hspace{-2.8cm}v_a\rightarrow v_{\a\adt}=(\s^a)_{\a\adt} v_a\ .
\ee
In {\eq{tor}} multiple  internal indices are antisymmetrised, while the spinor indices on $M$ and $\bar\Psi$ are symmetrised. The leading components of the Weyl multiplets are $\Psi, M$ and $\chi$ for $N=1,2,3$ respectively. For $N=4$ there is also a complex dimension-zero scalar field that does not appear directly in the components of the torsion tensor. Instead it parametrises the coset space $SU(1,1)/U(1)$. The $U(1)$ factor here is part of the superspace internal connection, but for $N=4$ not part of the gauge group of the superconformal group which is only $SU(4)$. An additional constraint is required at dimension one to enforce this. It is
\be
D_{\a i}\chi^{\a\, ijk}=0\ .
\la{}
\ee

{The components of the covariant curvature can easily be found; for example, at dimension one, they are  {\cite{Howe:1981gz}}:

\begin{align}
\label{cur}
\cR_{\a i \b j,\c \d}&=0\ ,\nn\w1
\cR_{\a i \b j,\cdt \ddt}&=-\ve_{\a\b}\bar{M}_{\cdt \ddt ij}\ ,\nn\w1
\cR _{\a i \b,k}{}^{j l}&=-\ve_{\a\b}\bar B_{ijk}{}^l\ ,\nn\w1
\cR_{\a i\bdt,k}^{\hspace{0.8em} j }{}^l&=-\bar\chi_{\bdt ikm} \chi_\a^{jlm}\ , 
\la{}
\end{align}
{ where
\be
B^{ijk}{}_l=\half D_{\a l} \chi^{\a ijk}\ .
\ee
}

{Further details of the torsions and curvatures, including the Schouten terms, can be found in \cite{Howe:1980sy,Howe:1981gz}.
} 
\section{$D=6$}
The $(1,0)$ theory in $D=6$ has been discussed in { $SU(2)$ superspace in \cite{Linch:2012zh} and in conformal} superspace   \cite{Butter:2016qkx}, \cite{Butter:2017jqu}, \cite{Butter:2018wss}. The theory was also discussed earlier in harmonic superspace in \cite{Sokatchev:1988nk} and, some years ago, in projective superspace \cite{Linch:2012zh}. In components, the $(2,0)$ supergravity dates back twenty years to \cite{Bergshoeff:1999db}. Below we summarise the local twistor formulation of  $(1,0)$ and $(2,0)$ $6D$ conformal supergravity recently presented in  \cite{Howe:2020xrg}.

The complex conjugate of a four-component $D=6$ spinor $u^\a$ is denoted $\bar{u}^{\adt}$ but this representation is equivalent to the undotted one as there is a matrix 
 $B_\a{}^{\adt}$ relating the two, $\bar u^{\adt}= \bar u^\a B_{\a}{}^{\adt}$. $B$ is unitary, $B^* B=1$, and satisfies $B\bar B=-1$.\footnote{We use the six-dimensional conventions of  \cite{Howe:1983fr}.} Similar remarks hold for the inequivalent spinor representation denoted by a lower index, $v_\a$ say. So a twistor $z$ consists of a pair of 4-component spinors and can be written
\be
z=\left(\begin{array}{c} u^\a \\ v_{\a} \end{array}\right)\ .
\la{3.1}
\ee

A supertwistor in $D=6$ can therefore be written in the form
\be
Z=\left(\begin{array}{c}
u^\a\\
v_{\a}\\
\l_i\\
\end{array}\right)\ .
\la{4.1}
\ee
where $i=1,\ldots 2N$ for $(N,0)$ supersymmetry, $N=1,2$. Here $(u,v)$ are commuting objects while $\l$ is odd. The superconformal group is $OSp(8|N)$ in complex superspace and preserves the orthosymplectic metric $K$, so for an element $g$ of the group we have
\be
g K g^{st}=K
=\left(\begin{array}{c c |c}
0&1_4&0\\
1_4 & 0& 0\\
\hline
0&0& J_0
\end{array}\right)\ .
\la{4.2}
\ee
where $^{st}$ denotes the supertranspose, which is the same as the ordinary transpose except for an additional minus sign for each element in the bottom left (odd) sector.
The $2N \xz 2N$ matrix $J_0$ is the $Sp(N)$ symplectic invariant. In real spacetime we need to impose the reality constraint
\be
g R g^*=R=\left(\begin{array}{c c |c}
0&B^{-1}&0\\
B & 0& 0\\
\hline
0&0& 1_{2N}
\end{array}\right)\ .
\la{4.3}
\ee
An element of the Lie superalgebra, $h$, has the form
\be
h=\left(\begin{array}{c c |c}
a^\a{}_\b& b^{\a\b}&\ve^{\a j}\\
c_{\a\b} & d_\a{}^\b& \vf_\a{}^j\\
\hline
\l_{i \b}&\r_i{}^{\b}& e_i{}^j
\end{array}\right)\ .
\la{4.4}
\ee
The orthosymplectic constraint implies that $b$ and $c$ are skew-symmetric and $d=-a^t$, as before, while
\be
e J_0=-J_0 e^t\ .
\la{4.5}
\ee
In indices, setting $(J_0)_{ij}=\h_{ij}$, this implies\footnote{Indices are raised and lowered according to the rule: $X^i=\h^{ij} X_j\Leftrightarrow X_i=X^j\h_{ji}$ with 
$\h^{ik}\h_{jk}=\d_j{}^i$.}
\be
e_{ij}:=e_i{}^k \h_{kj}=e_{ji}\ .
\la{4.6}
\ee
For the odd components we have
\be
\r=J_0 \ve^t \qquad   \l=J_0\vf^t \ .
\la{4.7}
\ee
or, in indices,
\begin{align}
\r_i{}^\b&=\h_{ij}\ve^{\b j}\Rightarrow \r_i{}^\b=-\ve^\b{}_i\ ,\nn\w1
\l_{i\b}&=\h_{ij}\vf_\b{}^j \Rightarrow \l_{i\b} = -\vf_{\b i}\ .
\la{4.8}
\end{align}
Next we need to impose reality in order to move to real superspace. This is done with equation (3.3) but this time with $R$ extended by the unit matrix in the odd-odd sector, as in \eq{4.3}.
The result of imposing $gRg^*=g$, at the Lie algebra level is that $a,b,c$ and $d$ obey the same conditions as in the bosonic case while $e$ satisfies 
\be
e=-e^*\ .
\la{4.9}
\ee
For the independent odd components of $h$ we have:
\begin{align}
\bar\ve^{\adt}_i&=-\h_{ij} \ve^{\b j} B_\b{}^{\adt}\ ,\nn\w1
\bar\vf_{\adt}^i&=(B^{-1})_{\adt}{}^\b \h^{ij}\vf_{\b j}\ .
\la{4.10}
\end{align}
These constraints simply mean that $\ve^{\a i}$ and $\vf_{\a i}$ are symplectic Majorana-Weyl spinors as one would expect. They are respectively the parameters for $Q$ and $S$ supersymmetry transformations.

The connection $\cA$ is
\be
\cA=\left(\begin{array}{c c |c}
\hat\O^\a{}_\b& iE^{\a\b}& E^{\a j}\\
iF_{\a\b} & \hat\O_\a{}^\b& F_\a{}^j\\
\hline
-F_{i \b}&-E_i{}^{\b}& \O_i{}^j
\end{array}\right)\ ,
\la{4.11}
\ee
where $E^A=(E^a,E^{\a i})$,  with $E^a=\half (\c^a)_{\a\b}E^{\a\b}$, will be identified with the even and odd super-vielbein one-forms of the underlying superspace, 
$F_A=(F_a,F_{\a i})$ is the connection for superconformal transformations, \ie $S$-supersymmetry and standard conformal transformations, $\hat\O^\a{}_\b$  ($\hat\O_\a{}^\b$) is the Lorentz plus scale connection and $\O_i{}^j$ the internal $\gs\gp(N)$ connection. On the bottom line, $F_{i\b}$ and $E_i{}^\b$ are transposes of $F_\a{}^j$ and $E^{\a j}$ with the internal index lowered by $(J_0)_{ij}=\h_{ij}$.
The curvature two-form, $\cF=d\cA+\cA^2$, has components  given in matrix form by:
\be
\cal{F}=\left(\begin{array}{c c |c}
\hat \cR^\a{}_{\b}\ &i\cT^{\a\b} & \cT^{\a j}\\
i \cS_{\a\b} & \hat{\cR}_{\a}{}^{\b}&\cS_{\a}{}^j\\
\hline
-\cS_{i\b}&-{\cT}_i{}^{\b}& \cR_i{}^j
\end{array}\right)\ ,
\la{4.12}
\ee
where, from (3.4), with $\a'\mapsto \a$, and with appropriate factors of $i$,
\begin{align}
\cT^{\a\b}&=\hat D E^{\a\b}+iE^{\a k} E_k{}^\b\ ,\nn\w1
\cT^{\a j}&=\hat D E^{\a j} + iE^{\a \c}F _\c{}^{j}\ , \nn\w1
\hat{\cR}^\a{}_\b&=\hat R^\a{}_\b - E^{\a\c} F_{\c\b} - E^{\a k} F_{k\b}\ ,\nn\w1
\cR_i{}^j&=R_i{}^j -F_{i\c} E^{\c j} -E_i^{\c}F_{\c}{}^j\ ,\nn\w1
\cS_{\a\b}&=\hat D F_{\a\b} +iF_{\a}{}^k F_{k\b}\ ,\nn\w1
\cS_{\a}{}^j&=\hat D F_{\a}{}^j +iF_{\a\c} E^{\c j}\ .
\label{4.13}
\end{align}
Here, $\hat D$ is the superspace covariant exterior derivative with respect to scale, Lorentz and internal symmetries, while the leading terms on the right, for the top four lines, are the standard superspace and torsion and curvature tensors for the corresponding connections (extended by the scale connection).

The detailed form of the Bianchi identity $\cD\cF=0$ is given, {\it mutatis mutandis}, by (3.6) to (3.8).  
 We shall now repeat the steps carried out in the non-supersymmetric case to reduce the conformal and superconformal boost parameters to derivatives of the scale parameter.
We introduce a group element $g(S,C,\C)$ where $S$ is a scale parameter and $\C_{\a}{}^{i}$ is an $S$-supersymmetry parameter. It is given by
\be
g=\left(\begin{array}{c c |c}
S^{-\half}& 0& 0\\
iS^{-\half} \tilde C& S^{\half}& \C\\
\hline
S^{-\half}J_0\C^t& 0& 1
\end{array}\right)\ ,
\la{4.14}
\ee
where the index structure is as above, in \eq{4.11} for example, where $J_0$ is the $Sp(N)$ invariant discussed previously, and where
\be
\tilde C +\tilde C^t +i \C J_0 \C^t=0\ .
\la{4.15}
\ee
If we write
\be
\tilde C=C-\frac{i}{2} \C J_0\C^t\ ,
\la{4.17}
\ee
then, from \eq{4.15}, $C$ is antisymmetric since $\C J_0 \C^t$ is symmetric.
Reality implies that
\begin{align}
C&=B C^* B\ ,\ 
\nn\w1
\bar\C&=-B^{-1}\C J_0\ .
\la{4.16}
\end{align}
Note also that the latter equation implies that $\C J_0\C=B(\C J_0\C)^*B$. 

Under such a transformation the components of $\cA$ transform as follows:

\begin{align}
\label{4.18}
E^{\a\b}&\mapsto S E^{\a\b}\ ,\nn\w1
E^{\a j}&\mapsto S^{\half}(E^{\a j} +i E^{\a\b} \C_\b{}^j)\ ,\nn\w1
F_{\a\b}&\mapsto S^{-1}\left(F_{\a\b} -\hat D C_{\a\b} -i(\hat D \C_{[\a}{}^k)\C_{\b]k} +2iF_{[\a}{}^k \C_{\b]k}+i\tilde C_{\c\a} E^{\c\d} \tilde C_{\d\b}-2\tilde C_{\c[\a} E^{\c k} \C_{\b]k}\right)\ ,\nn\w1
F_\a{}^j&\mapsto S^{-\half}\left(  F_\a{}^j-\hat D\C_\a{}^j  + i(E^{\b j} +iE^{\b\c} E_\c{}^j)\tilde C_{\b\a}-E^{\b k} \C_{\a k} \C_{\b}{}^j\right)\ , \nn\w1
\hat\O^\a{}_\b&\mapsto\hat\O^\a{}_\b - E^{\a\c} \tilde C_{\c\b}  - E^{\a k} \C_{\b k}+\half \d^\a_\b Y\ ,\nn\w1
\O_i{}^j&\mapsto \O_i{}^j +\C_{\a i} E^{\a j} +\C_{\a}{}^j E^{\a}_{\ i} + i\C_{\a i} E^{\a\b} \C_\b{}^j \ .
\la{4.18}
\end{align}

The curvature transformations are obtained from those for the potentials by replacing the latter by the former in the equations above. In addition, for the superconformal curvatures, the derivative terms in \eq{4.18} must be replaced by curvature terms as follows:
\begin{align}
\hat D C_{\a\b}&\mapsto  2\hat R_{[\a}{}^\c C_{|\c|\b]}\ ,\nn\w1
\hat D \C_\a{}^j&\mapsto R_\a{}^\b \C_\b{}^j + \C_{\a}{}^k R_k{}^j         \ .
\la{4.18.1}
\end{align}

If we take the trace of the third equation in \eq{4.13} we find that
\be
2\cR_0=2R_0- E^A F_A\ ,
\la{4.19}
\ee
where we have defined the super-vector-valued one-form $F_A=(F_a,F_{\a i})$. By adjusting this potential we can choose $\cR_0=0$ so that the (graded) antisymmetric part of $F_{AB}$ is now proportional to $R_{0 AB}$. Taking the trace of the transformation of $\hat\O^\a{}_\b$ we find
\be
2\O_0\mapsto 2\O_0 - E^a C_a -E^{\a i} \C_{\a i} + 2Y\ ,
\la{4.20}
\ee
so that we can use the parameters $C_a$ and $\C_{\a i}$ to set $\O_0=0$. This leaves residual transformations determined by the scale parameter $S$, 
\be
C_A=2 Y_A= 2S^{-1} D_A S\ ,
\la{4.21}
\ee
where $C_A=(C_a,\C_{\a i})$. We shall take the components of $Y$ to be given by $Y_A=(Y_a, \U_{\a i})$ in order to clearly distinguish the even and odd components where necessary.

The basic constraint that we shall choose is to set the even torsion two-form to zero,
\be
\cT^a=0\ ,
\la{4.13.5}
\ee
which is clearly covariant.
Using this, conventional constraints corresponding to connection choices (including superconformal ones) and the Bianchi identities, one finds that the covariant torsions (\ie the torsion components of $\cF$) are given by
\begin{align}
\cT_{\a i\,\b j}{}^{\c k}&=0\ ,\nn\w1
\cT_{a \b j}{}^{\c k}&=(\c_a)_{\b\d} G^{\c\d}_j{}^k:=(\c^{bc})_{\b}{}^{\c}G_{abc j}{}^k\ , \nn\w1
\cT_{ab}{}^{\c k}&=\Psi_{ab}{}^{\c k}\ ,
\label{4.13.6}
\end{align}
where $G_{abc jk}$ is anti-self-dual on $abc$ (by its definition), anti-symmetric on $jk$ and symplectic-traceless on $jk$ for $N=2$, and where $\Psi_{ab}{}^{\c k}$ is the gamma-traceless gravitino field strength. For the curvature tensor components we find
\begin{align}
\cR_{\a i \b j,kl}&=0\ , \nn\w1
\cR_{\a i\b j,cd}&=4i(\c^a)_{\a\b} G_{acd ij}\ , \nn\w1
\cR_{a\b j,cd}&= -\frac{i}{2}( \c_a\Psi_{bc}-\c_c\Psi_{ab} -\c_b\Psi_{ca})_{\b j}\ ,\nn\w1
\cR_{a\b j,kl}&=-8(\c_a \chi)_{\b (k,l)j}\ ,\nn\w1
\cR_{ab,cd}&=C_{ab,cd}\ ,\nn\w1
\cR_{ab,kl}&= F_{ab,kl}\ .
\label{4.13.7}
\end{align}
The dimension three-halves field $\chi^\a_{i,jk}$ is antisymmetric on $jk$  : it is a doublet for $N=1$ while for $N=2$ it is in the {\bf{16}} of $\gs\gp(2)$, \ie it is symplectic-traceless on any pair of indices. The graviton field-strength $C_{ab,cd}$ has the symmetries of the Weyl tensor, while $F_{ab,kl}$ is the $\gs\gp(N)$ field-strength tensor. We have thus located all of the components of the conformal supergravity field strength supermultiplets except for the dimension-two scalars which are given by
\be
C_{ij,kl}=D_{\a i} \chi^\a{}_{j,kl}\ .
\ee
for  the $(1,0)$ case this reduces to a singlet, while for $(2,0)$ $C_{ij,kl}$ is in the $14$-dimensional representation of $\gs\gp(2)$. { Fuller details of this multiplet can be found in  \cite{Howe:2020xrg}.}

The standard superspace torsion tensors differs from the covariant ones by components of the Schouten tensor $F_{AB}$.

\section{$D=5$}
The $D=5$ superconformal theory, which exists only for $N=1$, {where the algebra is ${\frak f}(4)$ \cite{Nahm:1977tg}}, { has been described in (conformal) superspace in {{\cite{Kuzenko:2008wr}}},
\cite{Butter:2014xxa}. However,}it does not fit into the supertwistor formalism as well as the other cases due to the constraints that one has to impose on the top left quadrant of $\cA$ (starting from the $D=6, (1,0)$ case). These are
\begin{align}
(\c^a)^\a{}_\b\hat\O^\b{}_\a&=0\ ,\nn\w1
\h_{\a\b} E^{\a\b}&=0\ ,\nn\w1
\h^{\a\b} F_{\a\b}&=0\ ,\nn\w1
(\c^a)_\a{}^\b \hat\O_\b{}^\a&=0\ ,
\la{5.1}
\end{align}
where $\a=1,\ldots 4$,  $i=1,2$ and where the vector indices are now five-dimensional. Similar constraints should hold for the curvature $\cF$, but the fermion bilinears in the $\cA^2$ term do not preserve them. Thus with the same definition of $\cF$ as before, $\cF=d\cA +\cA^2$, the curvatures in the top left quadrant should be defined by
\begin{align}
\hat\cR^\a{}_\b&=\cF^\a{}_\b-\frac{1}{4} (\c^a)^\a{}_\b (\c_a)^\c{}_\d \cF^\d{}_\c\ ,\nn\w1
\cT^{\a\b}&=\cF^{\a\b}-\frac{1}{4}\h^{\a\b} \h_{\c\d}\cF^{\c\d}\ ,\nn\w1
\cS_{\a\b}&=\cF_{\a\b}-\frac{1}{4}\h_{\a\b} \h^{\c\d}\cF_{\c\d}\ ,\nn\w1
\hat\cR_\a{}^\b&=\cF_\a{}^\b-\frac{1}{4} (\c^a)_\a{}^\b (\c_a)_\c{}^\d \cF_\d{}^\c\ ,
\la{5.2}
\end{align}
{ where the $\cF$s on the right-hand side are the top left quadrant entries in $\cF$,} and where $\h_{\a\b}$ is the $4\times 4$ symplectic matrix, { (the charge-conjugation matrix in $D=5$)}; the odd torsion, S-supersymmetry curvature and internal curvature are defined as before. The relations between the covariant curvatures and the standard superspace ones is the same as in \eq{4.13} with the difference that the unwanted terms have to be projected out as in the preceding equation. { For example, we have
\begin{align}
\label{5.3}
\cT^{\a\b}&=T^{\a\b}-i \langle E^{\a k} E^{\b}{}_k\rangle\ ,\nn\w1
\hat{\cR}^\a{}_\b&=\hat R^\a{}_\b - E^{\a\c} F_{\c\b} - \langle E^{\a k} F_{k\b}\rangle\ ,
\end{align}
where the angle brackets indicate that the symplectic trace terms on the first line and the $(\c^a)$-trace terms on the second line {have been projected out}. Note that the middle term on the right in the second equation is automatically $(\c^a)$-traceless because $E^{\a\c}$ and $F_{\c\b}$ are now five-dimensional vectors, so that multiplying by $\c^a$ and taking the trace is identically zero.}

We now briefly describe the constraints and the geometry that follows from them. As in the other cases we set $\cT^{\a\b}=0$ from which the even torsion two-form $T^{\a\b}$ takes its usual $\c$-matrix form, as can be seen from {the first line in \eq{5.3}} on setting the left-hand-side to zero. The only non-zero torsion components are at dimension one and three-halves. The former is
\be
\cT_{a \b j}{}^{\c k}=\d_j{}^k\left( (\c^{bc})_\b{}^\c G_{abc} +2 (\c^b)_\b{}^\c G_{ab}\right)\  ,
\la{5.4}
\ee
where $G_{abc}$ is the dual of $G_{ab}$,
\be
G_{abc}=\half \ve_{abcde} G^{de}\ .
\la{5.5}
\ee
The field $G_{ab}$ can be identified as the leading component of the $D=5,N=1$ Weyl multiplet. The covariant dimension three-halves torsion is the gamma-traceless gravitino field strength.  { Note that \eq{5.4} can be obtained by dimensional truncation from the $D=6, N=(1,0)$ expression.} The remaining components are a dimension three-halves spinor, $\chi_{\a i}$, and, at dimension two, the Weyl tensor $C_{ab,cd}$, the $\gs\gp(1)$ field strength $F_{ab,kl}$ and  a scalar field $C$.  The curvature components are formally the same as the $D=6, (1,0)$ case given in \eq{4.13.7}, except that
\be
\cR_{\a i\b j,cd}=4i\ve_{ij} \left( \h_{\a\b} G_{cd} +(\c^a)_{\a\b} G_{acd}\right)\ ,
\la{5.6}
\ee
while the dimension three-halves spinor $\chi_{\a i, jk}\rightarrow \ve_{i(j} \chi_{\a k)}$. Finally, the dimension two scalar can be defined to be
\be
C=D_{\a i}\chi^{\a i}\ .
\la{5.7}
\ee

{The standard superspace torsions and curvatures can be obtained from the covariant ones in the same way as in $D=6, (1,0)$, using \eq{4.13}. They are the leading terms on the right-hand sides of the first four equations, but with the symplectic and $\c^a$ traces projected out. The Schouten terms come from the equation for the odd torsion in \eq{4.13}.}
\section{Minimal approach}

We shall now describe the superspace geometry corresponding to these conformal supergravity multiplets from a minimal perspective. {Since the details coincide with those derived previously we will be brief.} We define a superconformal structure on a supermanifold with $(\rm{even}| \rm{odd})$ dimension $(D|D')$ to be a choice of odd tangent bundle $T_1$ (of dimension $(0|D')$ which is maximally non-integrable, so that the even tangent bundle $T_0$ is generated by commutators of sections of $T_1$, and such that the Frobenius tensor, $\bbF$, defined below, is invariant under 
$\bbR\oplus \gspin(1,D-1)\oplus\gg$, where $\gg$ is the internal symmetry algebra for the case in hand: $\gg=\gs\go(N)$ for $D=3$, $\gu(N)$ for $D=4$, $\gs\gp(N)$ for $D=6$, with $N=1,2$,  and for  $D=5$, with $N=1$.  The components of $\bbF$ with respect to local bases $E_{\a i}$ for $T_1$ and $E^a$ for $T_0^*$ are given by
\be
\bbF_{\a i \b j}{}^c= \langle[E_{\a i}, E_{\b j}], E^c\rangle=-i k_{ij} (\c^c)_{\a\b}\ ,
\la{6.8}
\ee
where $k_{ij}=\d_{ij}$ for $D=3$ and $\h_{ij}$ for $D=5,6$} respectively and where $\langle,\rangle$ denotes the pairing between vectors and forms. The  $\bbR$ factor denotes an infinitesimal scale transformation, $\d E^a=SE^a$, $\d E_{\a i}=-\half  S E_{\a i}$, while the spin and symplectic algebras act in the natural way on the spacetime and internal indices.  For $D=4$, $T_1$ is the sum of two complex conjugate bundles of dimension $2N$, $T_1=\cT_1\oplus \bar{\cT}_1$, and we have
\begin{align}
\bbF_{\a i \b j}{}^c&= \langle[E_{\a i}, E_{\b j}], E^c\rangle=0,\w1\nn 
\bbF_{\adt\,\bdt}^{\,i\ j\,c}&= \langle[\bar{E}_{\adt}^{i}, \bar{E}_{\bdt}^{j} ], E^c\rangle=0,\w1\nn 
\bbF_{\a i \bdt}^{\ \ \, j\,c} &= \langle[E_{\a i}, \bar{E}_{\bdt}^{j}], E^c\rangle=-i\d_i{}^j (\s^c)_{\a\bdt}\ .
\la{6.8}
\end{align}

We now introduce connections for $\gs\gp(N)$ and $\gspin(1,D-1)$ and define the torsion and curvatures in the usual way. Note that this procedure involves the complementary basis $E_{\a i}$ for $T_1^*$ which is only determined modulo $T_0^*$, i.e. shifts of the form
\be
E_{\a i}\mapsto E_{\a i} + L_{\a i}{}^b E_b\ .
\la{6.9}
\ee
We could include this in the structure group, along with a corresponding connection, but we shall instead follow the standard procedure of using this freedom to impose some additional constraints at dimension one-half. In addition we shall not include a scale connection so that we have the standard superspace geometrical set-up. 

{It is clear that the Frobenius tensor is invariant under scale transformations of the form $E^a\mapsto S E^a,\ E_{\a i}\mapsto S^{-\half} E_{\a i}$ (with the same transformation for $\bar E_{\adt}^i$ in $D=4$). If we impose constraints to determine $E_a$ and $E^{\a i}$ at dimension one-half we can then determine their transformations as well. A convenient one to consider, as it does not involve any connection terms, is 
\be
(\langle [E_{\a i}, E_b], E^c\rangle)_{(bc)}=0\ ,
\la{6.9.1}
\ee
where the symmetrisation is understood to include lowering the $c$ index on the left-hand side.} Making finite super-Weyl  (scale) transformations, we find that this constraint will be preserved if
\begin{align}
\label{6.9.2}
E_a&\mapsto S^{-1}\left( E_a +i (\c_a)^{\c\d} \U_{\c}{}^k E_{\d k}\right)\ ,\nn\w1
E^{a i}&\mapsto S^{\half} \left( E^{\a i}- iE^a(\c_c)^{\a\b} \U_\b{}^i\right)\  ,
\end{align}
where $\U_{\a i}=S^{-1} D_{\a i} S$. { The super-Weyl transformations for the Lorentz and internal connections are given in \eq{sw3} for $D=3$, \eq{sw4} for $D=4$ and \eq{4.18} for $D=6$. The $D=5$ case can be derived from $D=6$ by dimensional truncation.}

Identifying $\bbF_{\a i \b j}{}^c$ with the dimension-zero torsion $T_{\a i \b j}{}^c$, imposing suitable constraints on various components of the torsion corresponding to fixing the odd basis $E_{\a i}$ using \eq{6.9} and making appropriate choices for the $\gspin(1,D-1)$ and $\gg$ connections, one can show, with the aid of the usual superspace Bianchi identities and some algebra, that the components of the torsion and curvature tensors can be chosen to agree with those {derived previously. The finite super-Weyl transformations in the general case were given in \eq{}, from which it is straightforward to obtain particular cases.}

In addition to the fields of the conformal supergravity multiplet, this geometry will also contain the components of the super Schouten tensor $F_{AB}$, whose transformations can be found in \eq{4.18}. We can recover the covariant forms for the torsions and curvatures by reversing the steps made earlier.


\section{Local Super Grassmannians}

In the previous sections we have considered superconformal geometries starting from local supertwistors  in various dimensions. In all cases we have used the standard split of supertwistor spaces into even twistors together with additional odd components. However, we can choose different $(even|odd)$ splittings  which naturally give rise to formulations of superconformal geometry on different supermanifolds. These include chiral, projective \cite{Karlhede:1984vr}, \cite{Grundberg:1984xr},{\cite{Kuzenko:2009zu}}  and harmonic  superspaces \cite{Galperin:2007wpa},and can be thought of from the viewpoint of super flag manifolds as discussed from a pure mathematical perspective in {\cite{Roslyi:1986yg}},\cite{Manin}  and presented in a more accessible form in \cite{Howe:1995md}.  More recently the present authors have used them to discuss super-Laplacians and their symmetries in the context of rigid supersymmetry \cite{Howe:2015bdd,Howe:2016iqw,Howe:2018lwu}. In this section we shall give a brief discussion of some cases relevant to four-dimensional complex spacetime, deferring a fuller discussion to a later publication.

Grassmannians are spaces of $p$-dimensional planes in $\bbC^n$, with $p<n$, and super Grassmannians are spaces of $(p|q)$ planes in $\bbC^{m|n}$, with $p\leq m,\ q\leq n$ (but with $p+q< m+n$). For example, complex Minkowski space in $D=4$ can be thought of as an open set in the Grassmannian of 2-planes in $\bbC^4$,. In the super case, consider $D=4$-dimensional spacetime with $N=2$ supersymmetries. The corresponding supertwistor space is then $\bbC^{4|2}$, and  the Grassmannian of $(2|1)$ planes in $\bbC^{4|2}$ is known as analytic superspace. A supertwistor, \ie an element of $\bbC^{4|2}$ can be split into two halves $u^A$ and $v_{A'}$, where $A=(\a, 1)$ and $A'=(\a', 1')$, where we have used a prime to denote a dotted two-component spinor  index. This allows us to rewrite a supertwistor as
\be
Z^{\underline{A}}=\left(\begin{array}{c}
u^A\\
\ \ v_{A'}\ .
\end{array}\right)\ .
\ee
We can then define a superconformal connection $\cA$ acting on supertwistors in this basis in the form
\be
\cA=\left(\barr{cc} -\hat\O^A{}_B & E^{A B'}\\
F_{A' B}&\hat\O_{A'}{}^{B'}\earr\right)\ ,
\ee
with corresponding curvature
\be
\cF=d\cA + \cA^2=\left(\barr{cc} -\hat\cR^A{}_B & \cT^{A B'}\\
\cS_{A' B}&\hat\cR_{A'}{}^{B'}\earr\right)\ .
\ee
The action of the superconformal group on { $\cA$ and $\cF$} 
is formally the same as in the bosonic case, equation  {\eq{2.14}, but note that the diagonal transformations are not simply scale and Lorentz transformations, because the diagonal subgroups are also supergroups.  This is therefore a somewhat different way of looking at superconformal transformations which we propose to investigate further in a forthcoming paper. 

{Finally, we note that such geometries can also be thought of super versions of paraconformal geometries {\cite{BayEast}}.

{\section{Summary}

    In this article we have discussed the supergeometries describing off-shell conformal supergravity multiplets in $D=3$ for $N=1$ to $8$, in $D=4$ for $N=1$ to $4$, and in $D=6$ for $(N,0)$ supersymmetries with $N=1,2$, from the perspective of local supertwistors\footnote{ We note that there is a related construction in four dimensions described in \cite{Wolf:2007tx}. }. In this formalism one introduces connections taking their values in the superconformal algebras in the twistor representation, which can be thought of as an associated version of the Cartan connection formalism. { Similarly, the conformal superspace construction \cite{Butter:2009cp} is an adaption of the Cartan formalism to superspace,  and is expected to be equivalent to our local supertwistor approach.}. From this starting point one can then derive the standard superspace formalism in a systematic fashion. In order to specialise to the minimal off-shell conformal supergravity multiplets one then has to impose constraints. {  In addition, we discussed the $D=5, N=1$ case for which a slight amendment of the formalism is necessary}. We also showed that the same results can be obtained from the minimal formalism in which only the dimension-zero torsion, or Frobenius tensor, is specified.  This minimal formalism was previously applied to $D=3$ \cite{Howe:1995zm,Gran:2012mg}, where an additional self-duality constraint at dimension one is required,  while it
was also previously shown in the  $D=4$ case that the supergeometries also follow from the dimension-zero torsion constraint \cite{Howe:1980sy}. In this case an additional dimension one constraint is required to ensure the correct number of component fields. In section 10 we briefly introduced the idea of local super Grassmannians. Such superspaces are best viewed in terms of different $(even|odd)$ splittings of supertwistor space.}

\vskip 1cm

\noindent{\bf Acknowledgement:}\\
{ We are grateful to Sergei Kuzenko and Gabriele Tartaglino-Mazzucchelli for careful reading and comments on the manuscript.}
The research of U.L. is supported in part by the 2236 Co-Funded Circulation Scheme2 (CoCirculation2) of T\"UBITAK (Project No:120C067)\footnote{However the entire responsibility of the publication is  the authors'. The financial support received from T\"UBITAK does not mean that the content of the publication is approved in a scientific sense by T\"UBITAK.} as well as by L\"angmanska Kulturfonden.

\end{document}